\documentclass[11pt,twoside]{article}  
\usepackage{apn3conf}

\newcommand{\cpd}{CPD--56$^{\rm o}$8032}

\begin{document}   

%
%
%
%

\title{The Dual Dust Chemistry - Binarity Connection}

%
%
%

\author{Orsola De Marco}
\affil{American Museum of Natural Histroy,
    Central Park West at 79$^{\rm th}$ Street, New York, NY 10024}
\author{A.F. Jones}
\affil{Carter Observatory, New Zealand}
\author{M.J. Barlow}
\affil{University College London, UK}
\author{M. Cohen}
\affil{University of California at Berkeley, USA}
\author{H.E. Bond}
\affil{Space Telescope Science Institute, USA}
\author{D. Harmer}
\affil{National Optical Astronomy Observatories, USA}

%
%

\contact{Orsola De Marco}
\email{orsola@amnh.org}

%
%
%
%
%

\paindex{De Marco, O. }
\aindex{Barlow, M. J.}     
\aindex{Cohen, M.}     
\aindex{Bond, H. E.}     
\aindex{Harmer, D.}     

%
%

\authormark{De Marco, Barlow, Cohen, Bond \& Harmer}

%
%



\begin{abstract}          
Accumulating evidence points to 
a binary nature for the Wolf-Rayet ([WC]) central stars, a group that constitutes
about 15\% of all central stars of planetary nebula. From ISO observations,
a dual dust chemistry (oxygen- and carbon-rich) has been shown to be
almost exclusively associated with [WC] central stars, a fact that could be
explained by O-rich dust residing in a disk, while the C-rich dust being more
widely distributes. HST/STIS space resolved spectroscopy 
of the [WC10] central star \cpd, is interpreted as revealing a dust disk or
torus around the central star. This, together with \cpd's variable lightcurve
is taken as an indirect indication of binarity. Finally, we present here, for 
the first time, preliminary results from a radial velocity survey of central
stars. Out of 18 stars with excellent data at least 8 are radial velocity variables.
If these turn out to be binaries, it is likely that the central star 
binary fraction is as high as $\sim$50\%.  

\end{abstract}

%
%

\section{Prologue}
WC Wolf-Rayet central stars of planetary nebulae ([WC] central stars of PNe) are
H-deficient stars that exhibit strong ionic emission lines of helium,
carbon and oxygen from their dense stellar winds. Amongst the coolest
central stars in this group are \cpd\ (the nucleus of the PN He~3-1333)
and the central star of He~2-113 (both classified as [WC10], Crowther et al. 1998).
Cohen et al. (1986) found the mid-infrared KAO spectra of both these
objects to show very strong unidentified infrared bands (UIBs -- usually
attributed to polycyclic aromatic hydrocarbons, PAHs). Indeed, both the
nebular C/O ratios (De Marco, Barlow \& Storey 1997) and the ratio of UIB
luminosity to total IR dust luminosity (Cohen et al. 1989) for these two
objects are the largest known. It was therefore a major surprise
when mid- and far-IR Infrared Space Observatory spectra of these two
objects showed the presence of many emission features longwards of
20~$\mu$m that could be attributed to crystalline silicate and water ice
particles (Barlow 1997, Waters et al.  1998a, Cohen et al. 1999),
indicating a dual dust chemistry, i.e.  the simultaneous presence of both
C-rich dust and O-rich dust. The dual dust chemistry phenomenon in PNe
appears to show a strong correlation with the presence of a late WC
([WCL]) nucleus -- four out of six [WC8-11] nuclei studied by Cohen et al.
(2002) showed similar dual dust chemistries. In the context of a single star
scenario, this would point to a recent transition (within the last
$\sim$1000~yr) between the O- and the C-rich surface chemistries. However,
the probability of finding a post-AGB object that had recently changed
from an O-rich to a C-rich surface chemistry due to a third dredge-up
event should be very low indeed. An alternative scenario envisages these
systems as binaries (Waters et al. 1998a, Cohen et al. 1999, 2002), in
which the O-rich silicates are trapped in a disk as a result of a past
mass transfer event, with the C-rich particles being more widely
distributed in the nebula as a result of recent ejections of C-rich
material by the nucleus. 

\section{A Dusty Disk around \cpd }

\cpd\ exhibits a variable optical lightcurve (Pollacco et al. 1992, 
Jones et al. 1999), with minima which are $\sim$1~mag deep and have 
a 5-year pseudo-period (Cohen et al. 2002). This was 
interpreted by Cohen et al. as precession of a dusty disk. 
We present an updated light-curve for \cpd\ in Fig.~1.

De Marco et al. (2002) presented 
HST/STIS spatially-resolved spectroscopy of \cpd, obtained during its third visual 
light minimum, which revealed the
stellar near UV and blue optical continuum to be split into two spectra separated
by about 0.10 arcsec. Once other possibilities were excluded (e.g., instrumental problems 
or the presence of two stars), the simpler explanation is that \cpd's light is seen reflected 
by the upper and lower rim of a dusty disk or torus seen close to edge-on. If so, \cpd\
could be similar to the Red Rectangle (HD44179; Osterbart et al. 1997), 
whose binary central star light is seen indirectly,
reflected by the rims of an edge-on disk. In this binary scenario, \cpd's light-curve
could also be interpreted as \cpd's orbital motion taking
it in and out of alignment with a denser part of the disk, such that its brightness is 
modulated in step with the binary period. An alternative scenario that is more
in line with the asymmetric lightcurve declines, conceives a dust clump 
in orbit with the disk
passing in front of \cpd\ every 5 years. 
The companion cannot be bright enough to contaminate substantially its
spectrum, which is reasonably fit with a single 
Wolf-Rayet model atmosphere (De Marco \& Crowther
1998). \cpd's putative companion could plausibly be a low mass main sequence star, 
such as a K dwarf.  

Interestingly, the Red Rectangle, also a binary, is the only system to have a double 
dust chemistry and a central star which is {\it not} hydrogen-deficient. 
From this we could suggest that the Red Rectangle's A supergiant central star
(the other is thought to be an unseen white dwarf [Men'shchikov et al. 2002]) might
be on its way to becoming hydrogen deficient.

The conjecture that [WC] stars are binary systems is in harmony with the fact that
binarity is generally known to promote mass-loss. For instance, all the 
massive Wolf-Rayet stars in the metal-poor Small Magellanic Cloud are binaries.
This has been explained with the fact that the low metallicity of the SMC leaves
its stars with atmospheres with relatively low opacities, too low to develop 
the dense Wolf-Rayet winds. Hence the only massive Wolf-Rayet stars possible in the
SMC are those where a companion has facilitated mass-loss.

A direct detection of binarity via radial velocity monitoring (see Section 3), 
is unlikely
in the case of [WC] central stars, because of their intrinsically variable (e.g. Balick et al.
1996), broad emission lines.

\begin{figure}
\epsscale{0.6}
\plotone{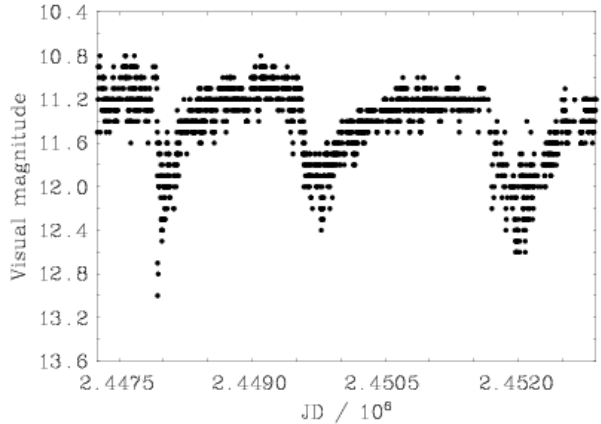}
\caption{The visual lightcurve of \cpd\ between March 1988 and September 
2003.} \label{fig:lightcurve}
\end{figure}

\section{Preliminary Results of a Central Star Radial Velocity Survey}

There are increasing indications that binary-star processes are intimately
related to the ejection of many, or possibly even most, PNe.  
The evidence includes: the fact that $\sim$10\% of PN nuclei are found
to be very close binaries (periods of hours to a few days; Bond \& Livio 1990,
Bond 2000) through photometric
monitoring; population-synthesis studies suggesting that these may be just the
short-period tail of a much larger binary population extending up to orbital
periods of several months (Yungelson et al. 1993, Han et al. 1995); 
and the prevalence of highly non-spherical
morphologies among PNe.

The photometric search technique does not work for binaries with periods
of more than a few days, since it relies on proximity effects. We have therefore
carried out radial-velocity monitoring of a sample of 
PN nuclei, in order to search for the anticipated population of binaries
with periods up to a few months.  If they do exist, there will be new
implications for the evolution of binary populations, the origin of compact
binaries (CVs, SN~Ia progenitors), and even the question of whether single stars
can produce visible ionized PNe at all.

This program was started in May 2002 at the 3.5-m WIYN telescope. 
Despite being plagued by very bad
weather and some instrument problems, we report that 8 of the 18 central stars for
which we have more than five data points have {\it clear} radial velocity variability. 
The independent, but similar program of Pollacco et al. (these proceedings) returned
a similar radial velocity variable central star fraction. 
However, it will take additional
observations for us to have robust statistics, and to determine orbital periods
needed to show conclusively that the velocity variations are due to binary
motion and not, for instance, to wind variability.

\acknowledgments
                                                                                                            
OD gratefully acknowledges Janet Jeppson Asimov
for financial support. OD and MC acknowledge support from
NASA grant HST-GO-08711.05-A. We are grateful to the 
WIYN telescope team.

%
%
%
%


\end{document}